# Temperature of the Source Plasma in Gradual Solar Energetic Particle Events

## Donald V. Reames

Institute for Physical Science and Technology, University of Maryland, College Park, MD 20742-2431 USA, email: dvreames@umd.edu

**Abstract** Scattering, during interplanetary transport of particles during large, "gradual" solar energetic-particle (SEP) events, can cause element abundance enhancements or suppressions that depend upon the mass-to-charge ratio $A/Q$ of the ions as an increasing function early in events and a decreasing function of the residual scattered ions later. Since the $Q$ values for the ions depends upon the source plasma temperature $T$, best fits of the power-law dependence of enhancements *vs.* $A/Q$ can determine $T$. These fits provide a fundamentally new method to determine the most probable value of $T$ for these events in the energy region 3–10 MeV amu$^{-1}$. Complicated variations in the grouping of element enhancements or suppressions match similar variations in $A/Q$ at the best-fit temperature. We find that fits to the times of increasing and decreasing powers give similar values of $T$, in the range of 0.8–1.6 MK for 69% of events, consistent with the acceleration of ambient coronal plasma by shock waves driven out from the Sun by coronal mass ejections (CMEs). However, 24% of the SEP events studied showed plasma of 2.5–3.2 MK, typical of that previously determined for the smaller impulsive SEP events; these particles may be reaccelerated preferentially by quasi-perpendicular shock waves that require a high injection threshold that the impulsive-event ions exceed or simply by high intensities of impulsive suprathermal ions at the shock. The source-temperature distribution of ten higher-energy ground-level events (GLEs) in the sample is similar to that of the other gradual events, at least for SEPs in the energy-range of 3–10 MeV amu$^{-1}$. Some events show evidence that a portion of the ions may have been further stripped of electrons prior to shock acceleration; such events are smaller and tend to cluster late in the solar cycle.

# 1. Introduction

The relative abundances of the chemical elements in energetic particles can experience considerable fractionation during their acceleration and transport from the many sources of astrophysical interest that we observe (*e.g.* Reames 1999). Sometimes old abundance observations yield new information when we learn to interpret the variations differently. Thus, source ionization-state dependence of abundance enhancements of elements in "impulsive" solar energetic-particle (SEP) events have been used recently to provide source plasma temperatures for SEP events acquired during 20 years of observations by the *Wind* spacecraft (Reames, Cliver, and Kahler 2014a, b, 2015). Since the impulsive acceleration process produces abundance enhancements that are strong functional power-laws of the mass-to-charge ratio $A/Q$ of each ion, the pattern of enhancements relates directly to the pattern of $A/Q(T)$, which depends upon the temperature $T$ at the time of acceleration. Can we use this powerful new technique to study the source plasma temperatures of the large "gradual" SEP events? Here the transit from a shock-acceleration source near the Sun to an observer near Earth involves a scattering mean-free-path $\lambda$ that, for constant particle speed, can vary as a function of $A/Q$, producing enhancement or suppression of the ions.

The distinction between gradual and impulsive SEP events has an extensive history (see reviews by Gosling 1993; Lee 1997; Reames 1999, 2013. 2015; Mason 2007). Impulsive SEP events are small, have relatively short durations, can have 1000-fold enhancements in $^3He/^4He$ (Mason 2007) and in heavy elements (Z>50)/O (Reames 2000, 2015; Mason *et al.* 2004; Reames and Ng 2004; Reames, Cliver, and Kahler 2014a), relative to the corona or solar wind, and are associated with solar flares or jets and type III radio bursts (Reames and Stone 1986). Ion acceleration appears to occur in regions of magnetic reconnection (*e.g.* Drake *et al.* 2009) on field lines that are open to interplanetary space with $^3He$ enhancements coming from wave-particle interactions (Temerin and Roth 1992; Roth and Temerin 1997; Liu, Petrosian, and Mason 2006).

In contrast, gradual SEP events would be better described as long-duration events. They are large, intense events, sometimes even producing ground-level events (GLEs) where GeV protons produce a measurable nuclear cascade through the Earth's atmosphere (Gopalswamy *et al.* 2012). Gradual events have average ion abundances similar to those of the corona or solar wind (*e.g.* Reames 2014; Schmelz *et al.* 2012) that have been accelerated at shock waves driven out from the Sun by coronal mass ejections (CMEs). Since the review of Meyer (1985), it has been clear that the average abundances of elements in large gradual SEP events are closely related to the corresponding abundances of elements in the solar corona (Reames 1995, 1998, 2014, Cohen *et al.* 2007). In fact, the average SEP abundances, together with those of the solar wind, *etc.*, are used to determine the best estimates of coronal abundances (Schmelz *et al.* 2012).

A 96% correlation between gradual SEP events and CME-driven shock waves was established early by Kahler *et al.* (1984) and that association has persisted (Kahler, 1992 1994; Gopalswamy *et al.,* 2002; Cliver, Kahler, and Reames, 2004; Rouillard *et al*., 2011, 2012) and been supported by studies of onset timing (Tylka *et al.* 2003, Reames 2009, Tan *et al.* 2013), transport (Ng, Reames, and Tylka 2003, Reames and Ng 2010), *in situ* observations (Desai *et al.* 2003, 2004, 2006), electron observations (Cliver and Ling 2007, 2009; Tan *et al.* 2011; Wang *et al.* 2012), and shock acceleration theory (Lee, 1983, 2005; Ng and Reames, 2008; Sandroos and Vainio, 2009).

After Mason, Mazur, and Dwyer (1999) found five-fold enhancements of $^3$He/$^4$He in gradual events, where they were not expected, it became apparent that shock waves could re-accelerate residual suprathermal ions from impulsive SEP events. Tylka *et al.* (2005) and Tylka and Lee (2006) were able to explain sharp increases or decreases in Fe/C above 10 MeV/amu in otherwise similar gradual events by the preferential acceleration of higher-speed impulsive suprathermal ions, depending also upon the angle between the magnetic field and the shock normal. In this theory, quasi-perpendicular shock waves preferentially select pre-accelerated ions that are injected since these higher-speed ions can return to the shock more easily from downstream. Thus, the accelerated ion population is related to the seed population, and its energy spectrum, sampled by the shock, and reacceleration plays an important well-documented role (Kahler 2001; Gopalswamy *et al.* 2002, 2004; Desai *et al.* 2003, 2004, 2006, Tylka *et al.* 2005; Tylka and Lee 2006; Cliver 2006; Mewaldt *et al.* 2007; Li *et al.* 2012; Reames 2013). Reames, Cliver, and Kahler (2014a) selected impulsive SEP events by their high (×4) enhancements of Fe/C, but these authors noted that in some cases these SEPs may have been reaccelerated by an accompanying fast shock.

However, it is not the abundances averaged over many gradual SEP events which interest us here, but the *A/Q* dependence of abundances and its variation during individual SEP events. Using low-energy measurements of averaged values of *Q* by Luhn *et al.* (1984), Breneman and Stone (1985) first pointed out that in some events successive ion abundances increased with *A/Q* and in others they decreased with *A/Q*. A reliable indication of these variations is generally shown by corresponding variations in abundance ratios like Fe/C or Fe/O that involve elements that span a wider range of *A/Q*.

As particles stream out from the Sun along magnetic field lines, they are scattered by resonant magnetic fluctuations such as Alfvén waves, which can be amplified by the streaming particles themselves to increasing the scattering of particles following behind (*e.g.* Ng, Reames, and Tylka 2003; Reames and Ng 2010). If the spectrum of waves is a power law in frequency then the scattering mean free path will be a power law in particle rigidity. Comparing different ion species at the same velocity, their scattering will differ as a function of the *A/Q* ratio of the ions. Typically, Fe, with a larger value of *A/Q*, scatters less than O, so that Fe tends to be enhanced earlier while C or O are retarded and decrease Fe/C or Fe/O later in the same event. While this pattern is often observed, as we shall see, solar rotation, which warps the magnetic field into the Parker spiral, sweeps the early particles to the east, causing an east-west asymmetry. Thus, events from western sources on the Sun tend to be dominated by Fe/C enhancements while those with central and eastern sources tend to have a net Fe/C depletion.

According to diffusion theory, an abundance ratio like Fe/C has a power-law dependence upon $\lambda$, which has a power-law dependence upon *A/Q*, times a complex exponential dependence on time during an event (see Appendix A). When plotted *vs. A/Q* this complex expression is seen to be approximately power-law as shown in Appendix A. Therefore, we assume a power-law relationship between abundances and *A/Q* that makes this study more tractable. We will also show examples where complex patterns of abundance enhancements of elements correspond well with equally complex patterns of *A/Q* at the best-fit plasma temperatures, even early in SEP events.

The SEP abundances in this article were measured using the *Low Energy Matrix Telescope* (LEMT: von Rosenvinge *et al.*, 1995) onboard the *Wind* spacecraft which measures the elements He through about Pb in the energy region from about $2 - 20$ MeV amu$^{-1}$ with a geometry factor of 51 cm$^2$ sr, identifying and binning the major elements from He to Fe onboard at a rate up to about $10^4$ particles s$^{-1}$. Instrument resolution and onboard processing have been described elsewhere (Reames *et al.*, 1997; Reames, Ng, and Berdichevsky, 2001; Reames, 2000; Reames and Ng, 2004). Typical resolution of LEMT from He isotopes through Fe was shown by Reames *et al.* (1997) and by Reames (2014) and resolution of elements with $34 < Z < 82$ by Reames (2000). The LEMT response was calibrated with accelerator beams of C, O, Fe, Ag, and Au before launch (von Rosenvinge *et al.*, 1995).

Throughout the paper we use the term "enhancements" to mean observed abundance ratios, usually relative to oxygen, X/O, that are all divided by the corresponding average or "coronal" abundance ratio inferred from gradual SEP events and listed in Reames (2014). Thus abundances are always relative to these coronal values. Enhancements with values < 1 are referred to as abundance "suppressions" or "depressions."

## 2. Analysis

The analysis here follows the technique used by Reames, Cliver, and Kahler (2014b) for impulsive events. In order to follow the time dependence frequently seen during events, yet retain sufficient statistical samples, we study 8-hr averages (always beginning at 0, 8, or 16 UT) during each gradual SEP event. We select periods with abundances either increasing or decreasing with $A/Q$ as indicated by enhancements in Fe/C, or specifically, in Fe/C/0.288, where 0.288 is the "coronal" Fe/C ratio as measured by the SEP average value (Reames 2014). Note that event periods without significant enhancement or depression in abundances provide no information on $A/Q$ or source temperature, and must be omitted.

Figure 1 shows typical values of $A/Q$ vs. $T$ used for elements in the region we intend to study. Values of $Q$ vs. $T$ are obtained from Arnaud and Rothenflug (1985), Arnaud and Raymond (1992), and Mazzotta *et al.* (1998) up to Fe, and from Post *et al.* (1977) above Fe.

**Figure 1.** $A/Q$ is plotted as a function of the theoretical equilibrium temperature for elements that are named along each curve. Points are spaced every 0.1 unit of log $T$ from 5.9 to 6.9.

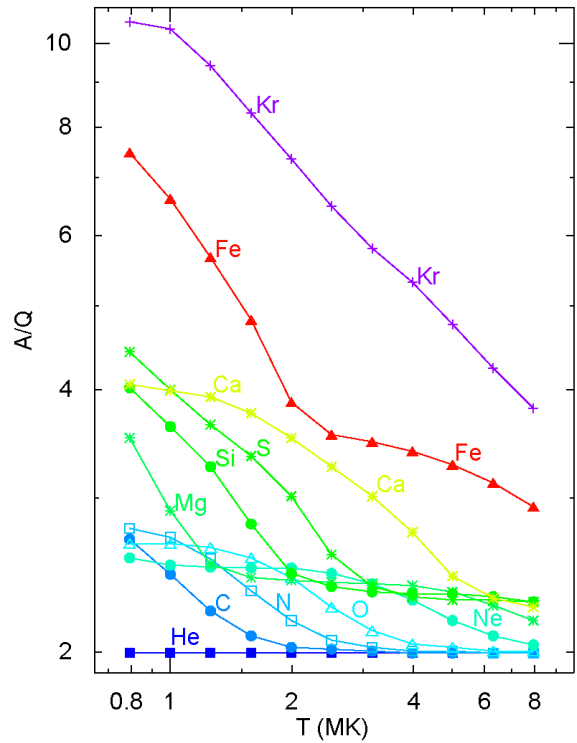

For each time period we calculate enhancements of the elements He, C, N, O, Ne, Mg, Si, S, Ar, Ca, and Fe, and the groups $34 \leq Z \leq 40$ and $50 \leq Z \leq 56$, although the latter groups contribute little statistically. Enhancements relative to O are determined at 3–5 MeV amu$^{-1}$, for most species, and are normalized to the SEP cor-onal values from Reames (2014). For elements that are poorly resolved at lower energies, such as Ar and Ca, enhancements relative to O at 5–10 MeV amu$^{-1}$ are used (see Reames 2014). The comparatively poorer quality of available H measurements have precluded their inclusion in this study. With a single ionization state, A/Q for H, like that for He, would be invariant throughout our temperature region, adding little to this study.

For each temperature point shown in Figure 1, a least-squares fit of enhancement *vs.* $A/Q(T)$ is obtained (best-fit examples are in the lower right panels of Figures 2, 4, 6, 7, and 9). These enhancements, relative to O, are normalized to coronal values as discussed previously. The fit, and the corresponding temperature, with the smallest value of $\chi^2$ *vs.* $T$ are selected as best fit (see upper right panels in Figures 2, 4, 6, 7, and 9). Gradual SEP events in the next section show the variety of their behavior.

## 3. Individual Events

Figure 2 shows an analysis of the large SEP event of 8 November 2000. The event has a source longitude of W75.

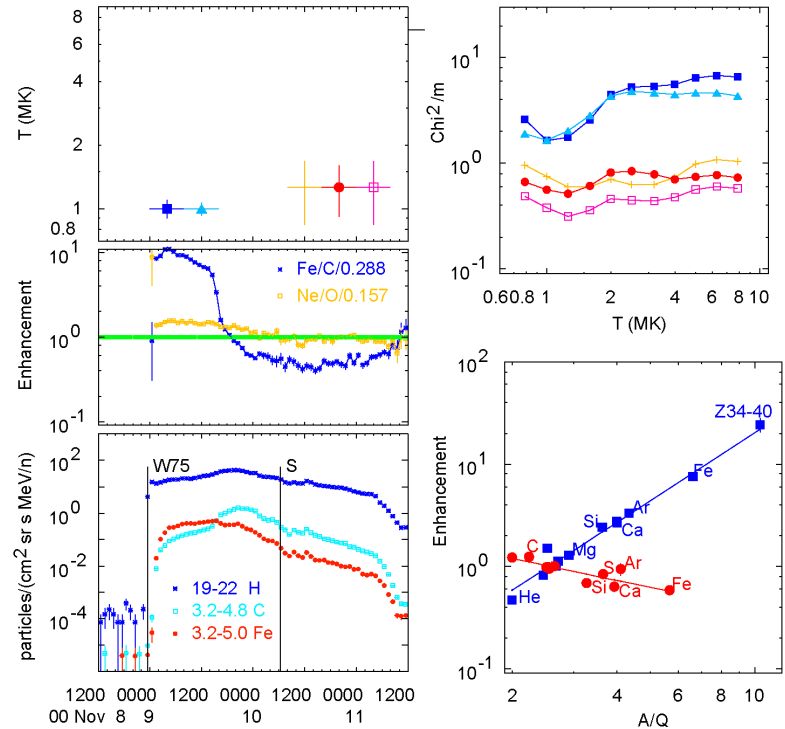

**Figure 2.** Clockwise from the lower left panel are the intensities of H, C, and Fe during the 8 November 2000 SEP event, the enhancements in Fe and Ne during the event, the best-fit temperatures in color-coded 8-hr intervals, values of $\chi^2/m$ *vs.* $T$ for each time interval, and best-fit enhancements, relative to O, *vs.* $A/Q$ and least-squares fits at two times.

Best-fit temperatures are shown for each 8-hr interval with symbols and colors that are also used for the corresponding plots of $\chi^2/m$ vs. $T$ and for two selected plots of enhancement *vs.* the best-fit *A/Q(T)*. We show $\chi^2/m$ where $m$ is the number of degrees of freedom, *i.e.* the number of enhancement points minus two (the number of fit constants: slope and intercept). When the fit is consistent with the errors, we should have $\chi^2/m \approx 1$. To account for non-statistical variations, we have included a 15% error, convolved with the statistical error, to obtain the weighting for each point in the fits (30% was the best value used for the impulsive SEP events as discussed by Reames, Cliver and Kahler 2015). These 15% errors are much smaller than the overall variations and might come from spatial variations in the source with better averaging in gradual SEP events that in impulsive. Also, the plasma may not be isothermal. Generally, the depth of the minimum in $\chi^2/m$ depends upon the steepness of the fitted line, *i.e.* large enhancements (or suppressions) of Fe/C provide well-defined fits and well-determined temperatures with small errors. Here, we use Fe/C rather than Fe/O to increase the leverage on the power law a bit. When enhancement *vs.* *A/Q* is flat, any temperature will do, and $\chi^2/m$ is also flat.

The two sampled plots of enhancement *vs.* *A/Q* in the lower right panel of Figure 1 show the quality of the best fits during times of enhanced and suppressed Fe/C. Note that:

      i) both early and late time intervals give similar temperatures, and

      ii) these are average coronal temperatures of $\approx$1 MK, *not* an impulsive-SEP temperatures of 2.5–3.2 MK.

While the lower right panel of Figure 2 shows that a power-law relationship between element enhancements and *A/Q* appear to be justified, this figure does not show the complex grouping of elements that have gone into this relationship. A comparison between element enhancements and temperature-dependent values of *A/Q* for the earliest 8-hour period in the 8 November 2000 SEP event is shown in Figure 3.

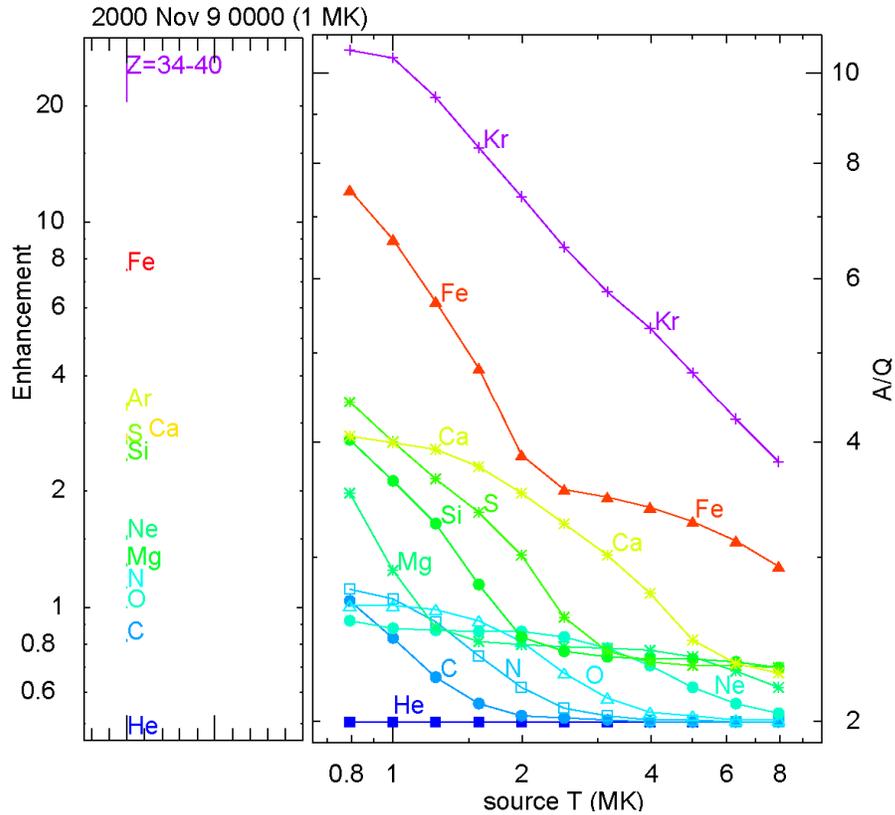

**Figure 3**. The left panel shows enhancements in element abundances during the interval 0000–0800 UT 9 November 2000. The right panel shows *A/Q vs. T* for various elements, as in Figure 1. The groupings of enhancements match those in *A/Q* near 1 MK.

Note in the enhancements in Figure 3, C, N, and O have moved up near Ne and Mg; near 1 MK, C, N, and O are not fully ionized like He, but have two orbital electrons. Meanwhile, Si has moved up, in both enhancement and *A/Q*, to join S, Ar and Ca. Spaced above this group is Fe and above it the measured group $34 \leq Z \leq 40$, represented by Kr in *A/Q*. In the plot of $\chi^2/m$ in the upper right panel of Figure 2, the filled blue squares show a minimum near one MK, indicating that no other temperature considered fits the observed pattern of enhancements as well. We must go down to one MK to adequately explain the original source plasma that was accelerated to produce the observed pattern of SEP element abundances. The strong correspondence between patterns of enhancements and those of *A/Q* suggests that our assumption of a power-law relationship did not seriously prevent us from deducing appropriate plasma temperatures near 1 MK. These complex enhancement patterns could not be produced by nonlinearities in the power-law relationship.

We find many events that show similar temperature behavior. Figure 4 shows an analysis of the event of 22 May 2013. This is another event from a western source longitude with periods of both enhanced and depleted Fe/C.

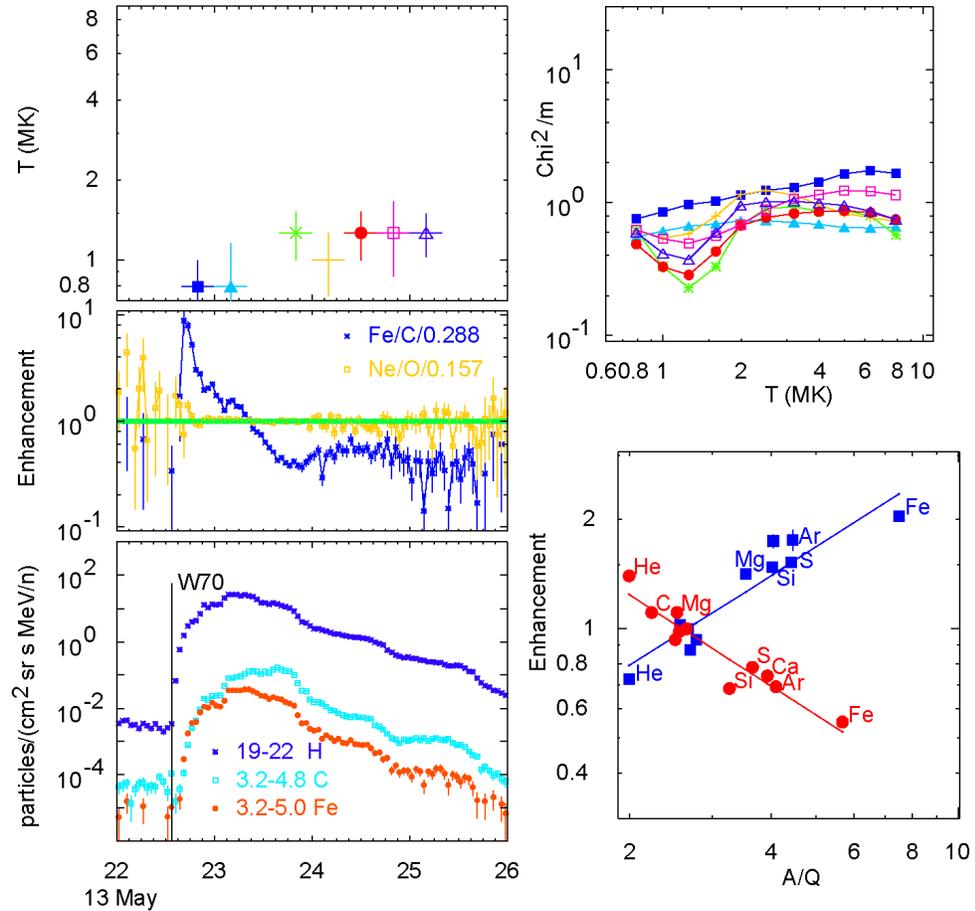

**Figure 4.** Analysis of the SEP event of 22 May 2013 as described for Figure 2.

The SEP event in Figure 4 has an extremely steep enhancement of Fe/C early in the event, yielding a best-fit temperature of 0.79 MK, at our lower boundary. Clearly these SEPs come from cool plasma, which measures 1.26 MK in the region of depressed Fe/C later on. In the lower right panel of Figure 4 notice that Mg and Si have similar enhancements and $A/Q$ values in the early period, but different enhancements and $A/Q$ values in the later period. The pattern of enhancements really does vary, over and above the variation in Fe/C.

In Figure 5 we compare the pattern of element enhancement for the first 8-hr period in the 22 May 2013 SEP event with the theoretical plot of $A/Q$ vs. $T$. We have extended the latter plot to slightly lower temperatures that are suggested by the best-fit plot of $\chi^2/m$ in Figure 4.

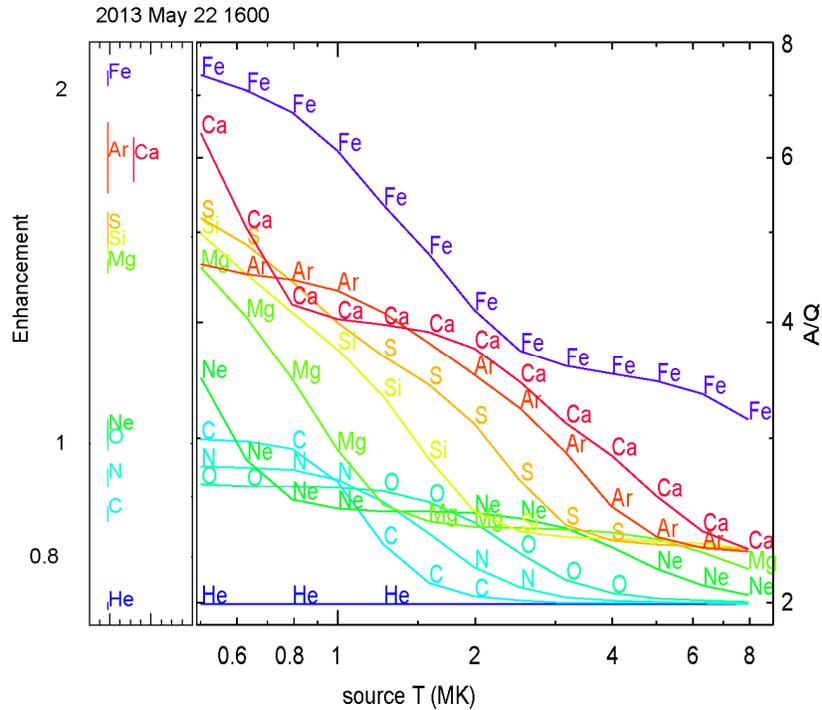

**Figure 5**. The left panel shows enhancements in element abundances during the interval 1600–2400 UT 22 May 2013. The right panel shows *A/Q vs. T* for various elements, as in Figure 1. The groupings of enhancements match those in *A/Q* near 0.6 MK.

In Figure 5, C, N, O, and Ne group well above He as expected for the low temperature where only He is fully ionized, although C and O are reversed from our expectations but within 15% errors. Mg and Si have moved well above Ne to join S, just below Ar and Ca. The spacing between S, Ca, and Fe seems appropriate for 0.6 MK. The enhancement of Ar is somewhat greater than expected, but its statistical error is large. The spacing between Ne and Mg is much too large for any temperature above 1 MK. A simple linear scaling between the log of the enhancement and log (*A/Q*) is required to produce the observed correspondences in Figure 5. As in the 9 November 1998 event shown in Figure 3, this event shows that a power-law approximation is appropriate and that we can measure source plasma temperatures in gradual SEP events using element abundances. These times occur early in SEP events when linear behavior is least likely, according to diffusion theory; at later times it is expected. Below we will show a third example with a higher source temperature.

Figure 6 shows an analysis of the SEP event of 24 August 1998. This event has a source at a solar longitude of E10 and shows only depressed values of Fe/C.

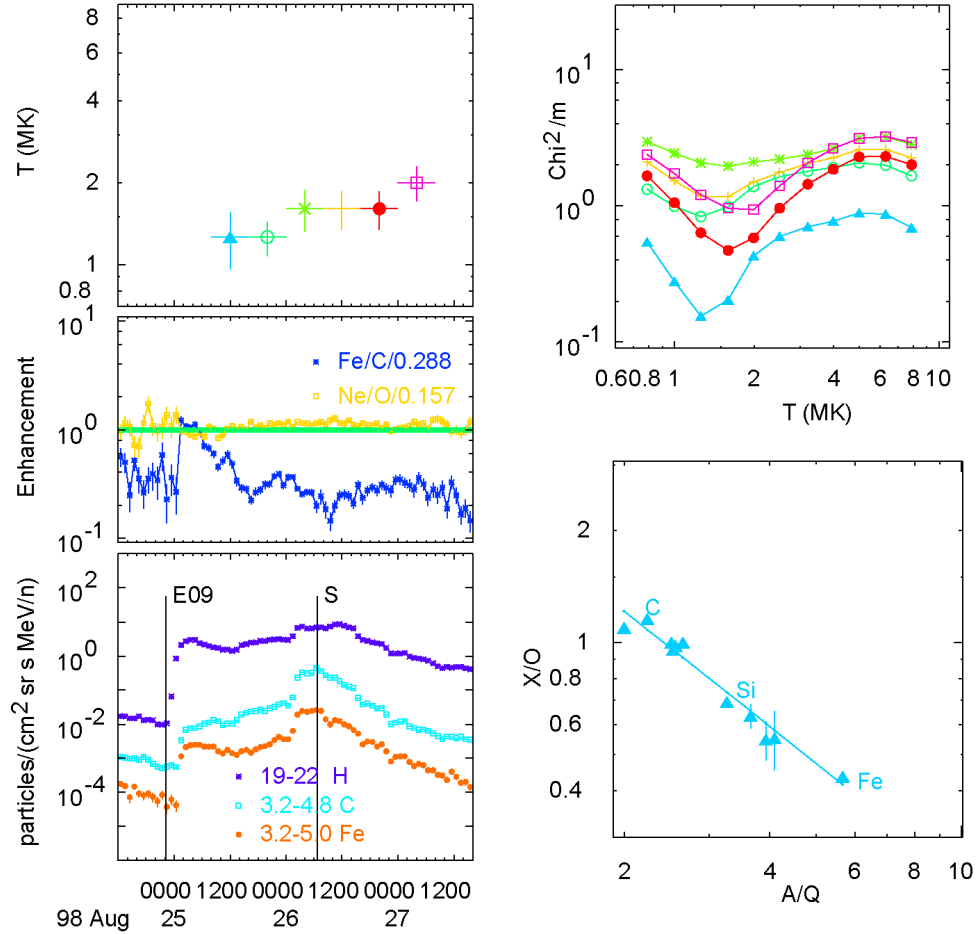

**Figure 6.** Analysis of the SEP event of 24 August 1998 from a source at E09 in panels as described for Figure 2

The intensities are smaller for this event so the statistical errors are larger especially for Ar and Ca enhancements. Nevertheless, we find a source plasma temperature varying from 1.3 MK to 2.0 MK during the event.

In the 24 August 1998 event the abundances are depressed so Figure 7 we compare the inverse of the depression of the abundances for the first 8-hr period beginning at 0800 on 25 August with the *A/Q vs. T* plot beginning with data for 1.26 MK.

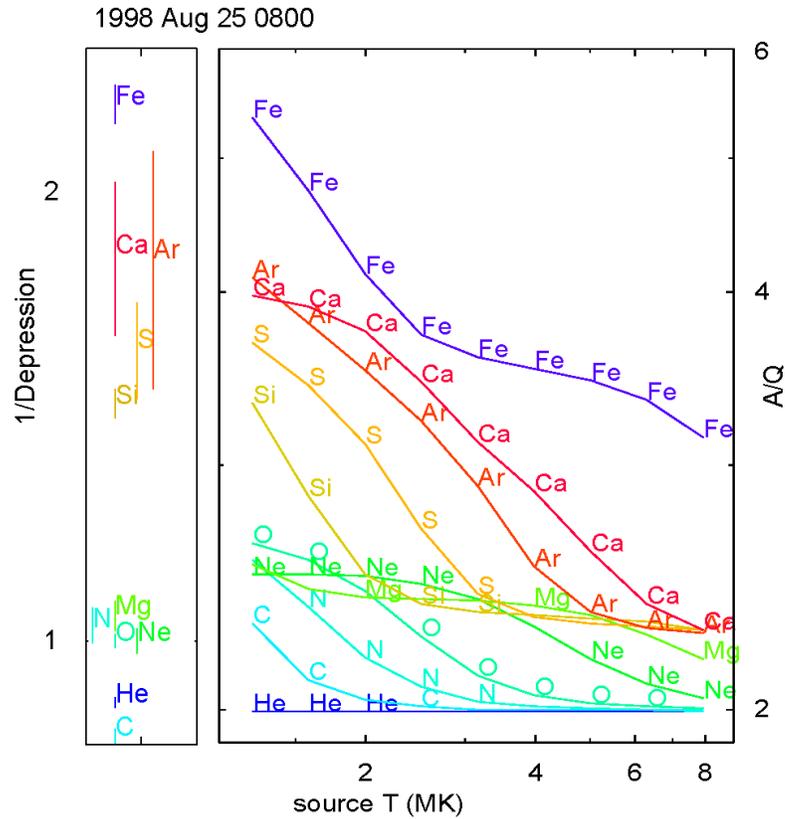

**Figure 7.** The inverse of the depression of the abundances during 0800–1600 UT on 25 August 1998 in the left panel is compared to the theoretical plot of *A/Q vs. T* in the right panel.

The pattern of abundances in Figure 7 shows He and C at the minimum with a grouping of N, O, Ne, and Mg above, typical of about 1.5 MK. Well separated above them are Si, S, Ar and Ca with large errors, and finally Fe. The patterns of abundance depressions map to patterns of *A/Q* showing approximate power-law behavior about as well as for the enhancements in other events.

Figure 8 shows an analysis of the gradual SEP event of 14 November 1998. This event shows source temperatures in the range of 2.5–3.2 MK typical of impulsive SEP events and probably involves reacceleration of material from an impulsive SEP event (Reames, Cliver, and Kahler 2014b). We will see that about 24% of the gradual events fall in this category.

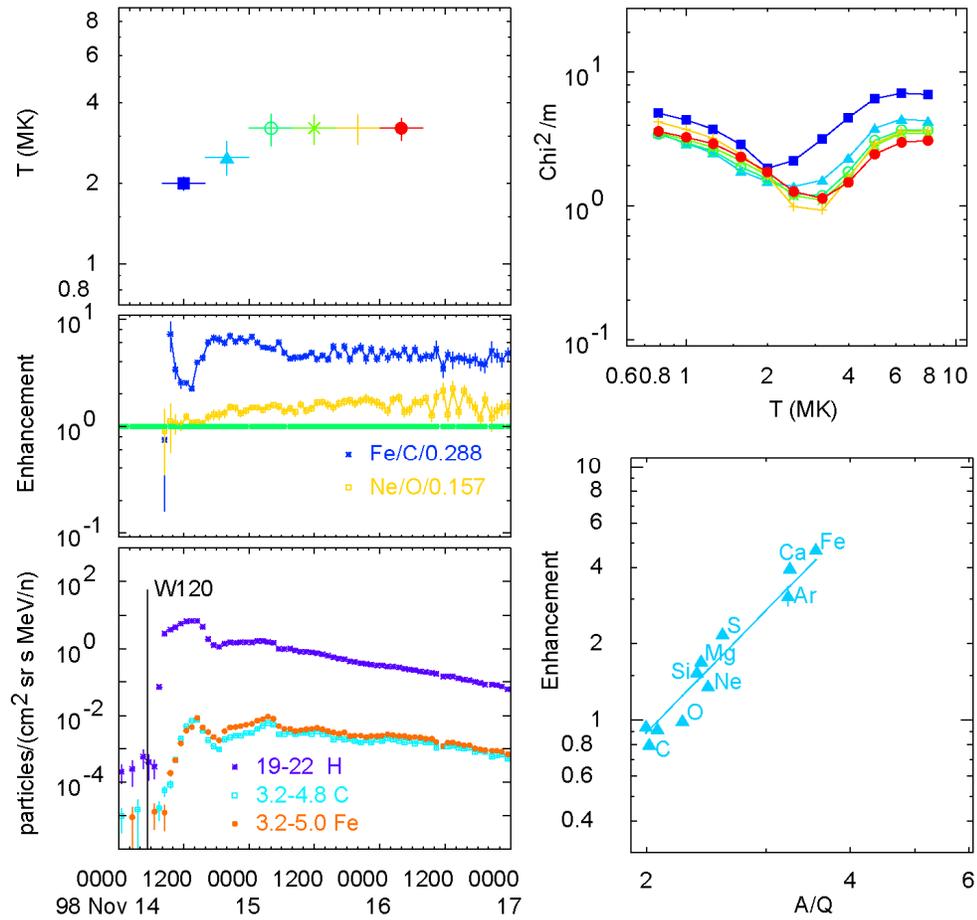

**Figure 8**. An analysis of the gradual SEP event of 14 November 1998 as described for Figure 2. This event shows source temperatures typical of material from an impulsive SEP event.

The lower right panel in Figure 8 samples the second 8-hr period in the event where the enhancement in Ne does not exceed those in Mg and Si. However, the center panel on the left shows the Ne/O enhancement rising strongly as the event progresses and the source temperature rises to 3.2 MK. We found previously that ×4 enhancements of Fe and ×2 enhancements of Ne typify impulsive SEP material (Reames, Cliver, and Kahler (2014a, b).

Since we have deduced a higher source plasma temperature for the event in Figure 8 than for the previous events, we compare the observed enhancements for the second interval with the relevant *A/Q* values in Figure 9.

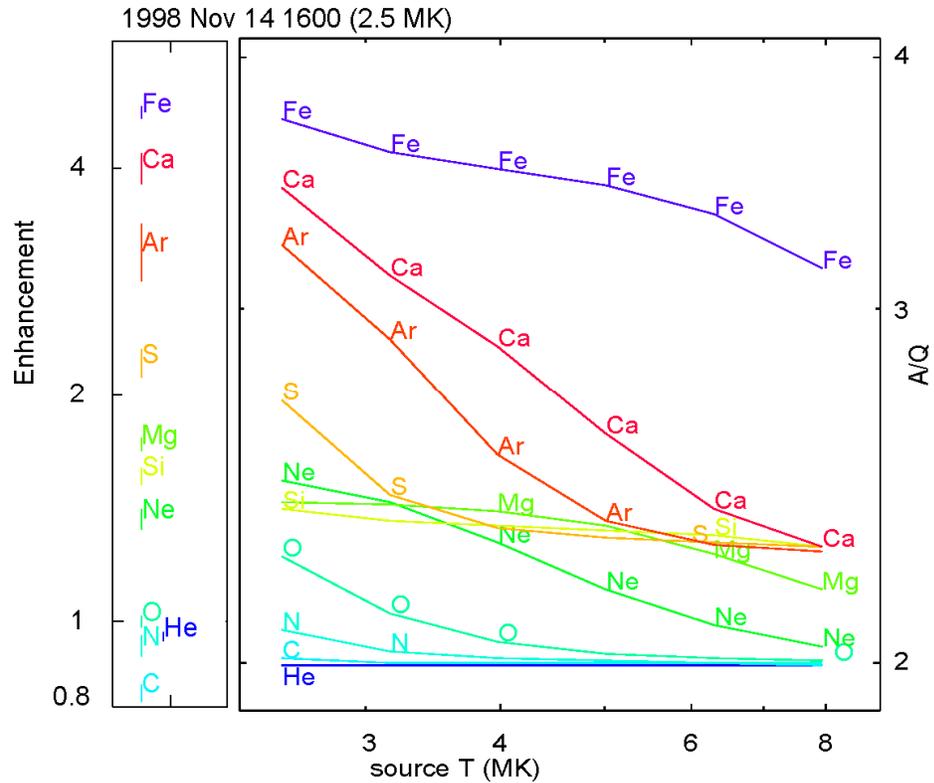

**Figure 9.** The left panel shows enhancements in element abundances during the interval 1600–2400 UT 14 November 1998. The right panel shows *A/Q vs. T* for various elements, as in Figure 8. The groupings of enhancements match those in *A/Q* near 2.5 MK. Compare with 1 MK patterns in Figure 3.

In the enhancements in Figure 9, C, N, and O have joined He; in fact C is even below He, probably reflecting the 15% errors we have assumed. He and C are fully ionized at 2.5 MK. Ne, Mg, and Si are grouped at higher enhancements; these elements have 2 orbital electrons at 2.5 MK. S, Ar, Ca, and Fe are spaced out at higher enhancements, as they are at higher *A/Q*. This pattern should be compared with that for 1 MK in Figure 3. It is the different pattern of groupings that determines the minimum value of $\chi^2$ and hence *T*. Again we find that the assumption of a power-law relation between enhancements and *A/Q* has lead to a reasonable abundance distribution and source temperature. In this case, however, the enhancements were probably already produced in the impulsive seed population rather than by transport from the shock source.

Finally, Figure 10 illustrates the event of 22 August 2005 with a different type of behavior altogether.

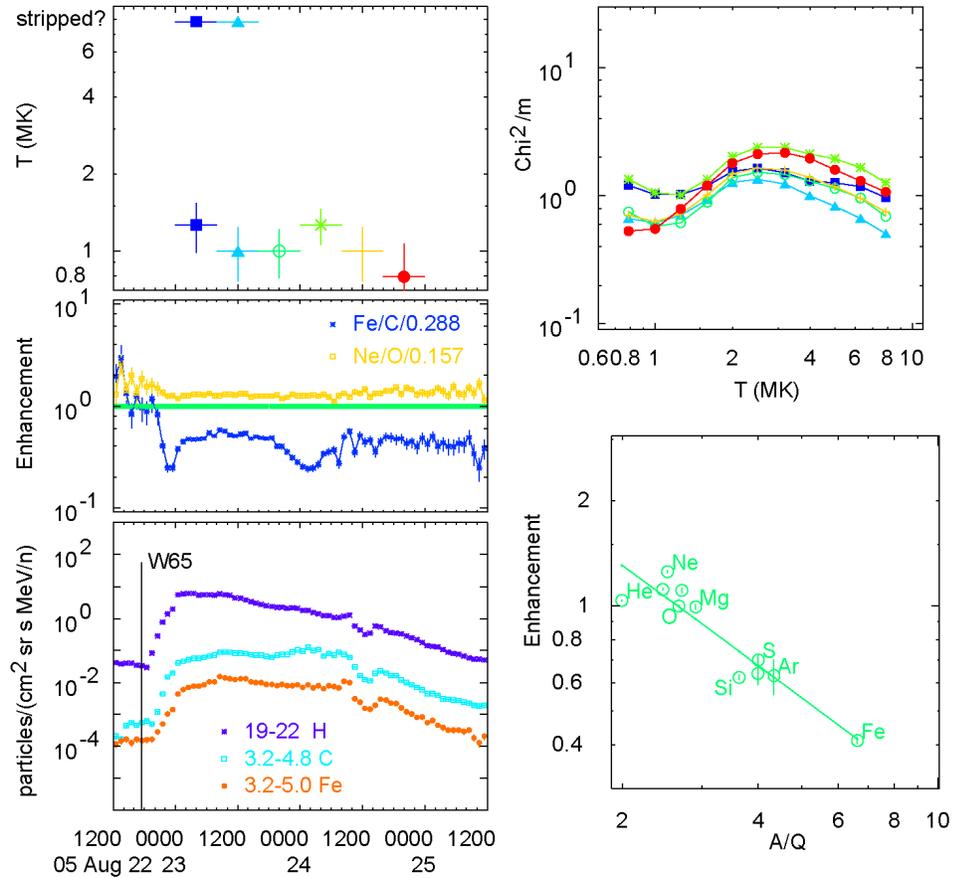

**Figure 10**. An analysis of the gradual SEP event of 22 August 2005 as described for Figure 2. Hot plasma or stripping?

The event in Figure 10 appears to show hot plasma early at the minimum in $\chi^2/m$, but also a strong local minimum at ≈1 MK, especially later. The high-temperature minimum is deeper early in the event.

At the highest temperature we consider, 7.9 MK, the elements up through Ne are fully ionized with $A/Q$=2, and $Q_{Fe}$= 19. Alternatively, this may be an equilibrium charge distribution seen when energetic ions from an earlier event have passed through a small amount of material in the low corona and have been stripped of some or all electrons prior to acceleration. Equilibrium ionization states that increase systematically with energy have been observed below ≈1 MeV/amu by DiFabio *et al.* (2008). This energy-dependent ionization can not be thermal but must result from passage of the ions through matter (Klecker *et al.* 2006).

The lower right panel of Figure 10 shows enhancements *vs. A*/Q for the third 8-hr period for *T*=1 MK. It is difficult to show the periods with *T*= 7.9 MK on the same scale since all values of *A/Q* are very small. For the data plotted, we can imagine all the elements up to Mg shifted to $A/Q$=2 since they all have small enhancements relative to He. In fact, the main unique feature of this event is that the observed He/O value is low for 1 MK but compatible with 7.9 MK.

The ambiguous SEP event in Figure 10 probably involves source material that contains both ≈1 MK and stripped ions. We only see the region of suppressed Fe/C for this event, where C is scattered more than Fe. The stripped ions with low $A/Q$ will be scattered more then the high-$A/Q$ ions from 1 MK and will be preferentially enhanced in this region.

## 4. Distributions

One goal for event selection in this study was to examine consistency of temperature measurements during long periods in gradual SEP events. Since gradual SEP events are primarily big proton events, we began with the criterion of Tylka *et al.* (2005) to select candidate events with >30 MeV proton fluence above $2 \times 10^5$ cm$^{-2}$ sr$^{-1}$, extending the time coverage to the 20-yr period from 4 November 1994 until 1 January 2015. However, a few (three) smaller events were also retained for comparison. During the events we examined 8-hr time periods that had sufficient increase or decrease in Fe/C to produce a positive or negative power of enhancement *vs. A/Q(T)* well above the errors, generally an absolute value of the power >0.5. Thus, time periods with >50% temperature errors were excluded. For this preliminary study, we were left with 45 events that had at least four useable 8-hr time periods available so we could examine the variation and consistency of the measured temperature. These 45 gradual SEP events are listed in Table B1 in Appendix B.

For these 45 events we determined the average or most probable temperature. All events like the one shown in Figure 10 showed a lower temperature, in addition to the 7.9 MK, at some times and showed a local minimum in $\chi^2/m$ at a lower temperature in all cases (the event in Figure 10 is an extreme example). For these events, we accepted the value at lower temperature, but separately noted evidence for the presence of hot or stripped ions.

From the general distribution of 45 gradual SEP events we found that:

i) 11 events (24%) showed source plasma temperatures of 2.5–3.2 MK similar to impulsive SEPs,

ii) 31 events (69%) had source plasma temperatures of ≤ 1.6 MK,

D. V. Reames

ii) 11 events (24%) show some evidence of hot plasma or stripped ions,

iv) all events showing stripped ions also have other source plasma with $T \leq 1.6$ MK, none with $T > 1.6$ MK

v) no events had actual average source plasma temperatures at or above 4 MK.

Other aspects of the event distribution are shown in Figure 11 where source temperatures are displayed as a function of time and of the >30-MeV proton fluence from the *Geostationary Operational Environmental Satellite* (GOES; http://www.ngdc.noaa.gov/stp/satellite/goes/dataaccess.html.)

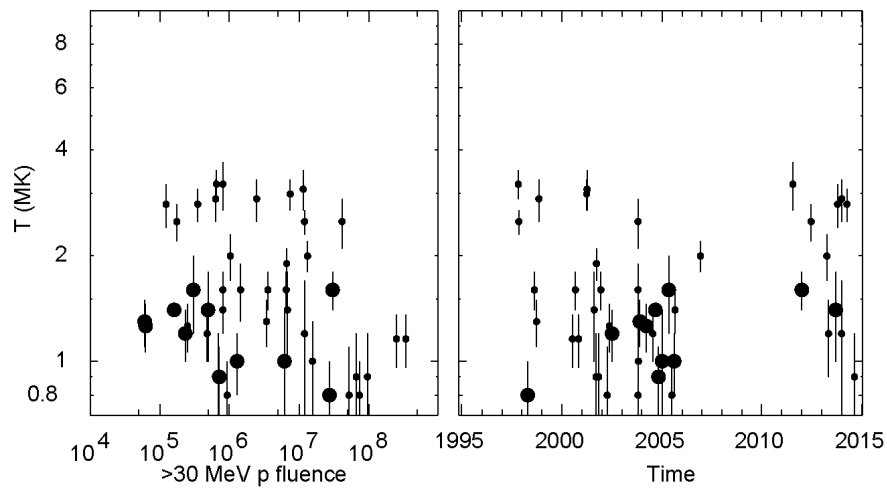

**Figure 11**. Event-averaged source-plasma temperatures are shown *vs.* the GOES >30 MeV proton fluence (left panel) and the event time (right panel) for gradual SEP events in this study. Events showing a presence of stripped ions have larger circles.

Since the correlation coefficient for the plot of log *T vs.* log fluence in the left panel of Figure 11 is only -0.26, there is only a hint that larger events have lower source plasma temperatures. However, it is true that events with stripped ions tend to be small (8 of 11 below $2 \times 10^6$ cm$^{-2}$ sr$^{-1}$). From the plot of log *T vs.* time in the right panel, we see that events with stripped ions also tend to cluster late in the solar cycle in 2004 and 2005. The lack of any events with a systematic source temperature with $T \geq 4$ MK is clear from the figure.

Finally, in Figure 12, we show the source-plasma temperature *vs.* CME speed derived from the Large Angle and Spectrometric Coronagraph (LASCO) on board the *Solar and Heliospheric Observatory* (SOHO) *via* the LASCO CME catalog (Gopalswamy *et al.,* 2009; http://cdaw.gsfc.nasa.gov/CME_list/). Acceleration of cooler plasma seems to require a faster CME.

**Figure 12**. The source-plasma temperature is shown as a function of the speed of the associated CME. The unweighted correlation coefficient for this plot is -0.49. Shocks of faster CMEs include more ambient coronal material.

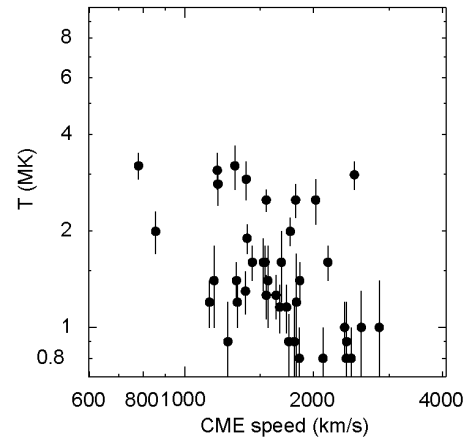

# 5. Coronal Abundances Revisited

The reference abundances, used to determine the abundance enhancements, were determined by averaging many over many gradual events (see Reames 1995, 2014). This assumed that abundance increases during one time period would be compensated by decreases during another period. If all the plasma were sampled from the corona, this might be the case. However, the presence of impulsive SEP material, always enhanced from the 2.5–3.2 MK source, would be expected to bias the average abundances toward Fe-rich material, enhanced in elements with high $A/Q$.

In order to assess the magnitude of this problem, we have determined the average abundances by excluding all time periods when $T$>1.5 MK. This excludes periods when significant impulsive-SEP material might be dominant and also periods with strong evidence of the presence of stripped material showing as 7.9 MK. As in previous studies (Reames 2014) we include an error, 5% in this case, on each 8-hr time period to keep a few intense periods with small statistical errors from dominating the average. The resulting element abundances, normalized to O=1000, are shown in Table 1. Errors in the table are errors in the mean.

**Table 1.** Reference abundances used and cool-plasma abundances.

| Element | Reference | T<1.5 MK |
|---------|-----------|----------|
| He | 57000±3000 | 56000±2000 |
| C | 420±10 | 426±6 |
| N | 128±8 | 131±2 |
| O | 1000±10 | 1000±7 |
| Ne | 157±10 | 169±4 |
| Mg | 178±4 | 166±5 |
| Si | 151±4 | 128±5 |
| S | 25±2 | 25±2 |
| Ar | 4.3±0.4 | 5.1±0.3 |
| Ca | 11±1 | 11.4±0.6 |
| Fe | 131±6 | 111±6 |

The table shows a 15% decrease in Fe/O, with similar decreases in Mg and Si. These changes are more likely to change the slope of the power-law fit *vs.* *A/Q* than the assigned temperatures. While the amplitude of the dependence of the abundances on the first ionization potential (FIP; see Reames 2014) may be reduced somewhat, we see no need to repeat the analysis of either the gradual or the impulsive SEP events with these new abundances since we have already assumed a 15% error in the abundances. As always, our ability to measure average coronal abundances depends upon having the correct weighting between Fe-rich and Fe-poor periods to balance the contributions of scattered and un-scattered ions.

# 6. Discussion

In quasi-perpendicular shock waves, the magnetic field direction $\theta_{Bn}$ is nearly orthogonal to the shock normal (>60°) and lies even nearer the plane of the shock downstream. Particles traveling along *B* downstream of the shock must have a speed exceeding $V_S \sec \theta_{Bn}$ to overtake the shock so as to scatter back and forth across the shock for acceleration. Tylka *et al.* (2005; see also Tylka and Lee 2006) suggested that this high threshold energy for acceleration at quasi-perpendicular shock waves would lead to preferential acceleration of residual pre-accelerated ions from impulsive SEP events. It is entirely possible that the 24% of gradual events with high source plasma temperatures of 2.5–3.2 MK were re-accelerated by quasi-perpendicular shock waves with high injection thresholds that required pre-accelerated impulsive suprathermal ions. However, calculations by Giacalone (2005) point out that turbulence near the shock can produce large values of $\delta B/B$ that will reduce the importance of $\theta_{Bn}$. Under these circumstances the gradual SEP events showing 2.5–3.2 MK plasma may be produced in regions where impulsive suprathermal ions dominate. Note that we do not find only a small component of impulsive suprathermal ions as seen by Mason, Mazur, and Dwyer (1999), but the dominant constituent. In addition, Desai *et al.* (2006) estimated that 75% of the ions below 1 MeV amu[-1] were from impulsive material.

Ten of our 45 events are GLEs. These events have a source temperature distribution that is similar to the other events at least for SEPs in the energy-range of 3–10 MeV amu[-1]; three GLEs are in the 2.5–3.2 MK interval. However, none of them show evidence of the presence of stripped ions. They are mainly distinguished by their high proton fluences, near or above $10^7$ cm[-2] sr[-1]

It is surely significant that 24% of the gradual SEP events show source plasma characteristic of impulsive SEP events, yet they have long durations, have high proton fluences, three are GLEs with GeV protons, and all but one have associated CMEs speeds above 1000 km s[-1]. If we aspire to distinguish gradual and impulsive events by their dominant acceleration physics, there is little doubt that these are shock-accelerated SEPs, just like the 69% of events with ambient coronal source plasma, even though the seed population is dominated by a remnant impulsive population of ions pre-accelerated in magnetic reconnection, suggesting preferential acceleration at select locations or by a quasi-perpendicular shock.

Actually the seed population of suprathermal ions available for shock acceleration may be quite complex. When we consider discrete events we tend to think of competition between coronal plasma and impulsive suprathermals which may be preferentially selected depending upon $\theta_{Bn}$ (Tylka and Lee 2006). However, there may also be suprathermal ions from earlier gradual events, which already involved a mixture of sources. As shocks move out from the Sun they reaccelerate material accelerated earlier (*e.g.* Desai *et al.* 2006) that may have undergone fractionation during transport or stripping, if it came from the low corona.

Most of our gradual SEP events (69%) do *not* have significant access to pre-accelerated impulsive ions – either these ions are not preferentially selected or they are not present at all. These events, with equally fast CME-driven shock waves, succeed in accelerating to high energies, ambient coronal plasma of $\approx 1$ MK. These events, dominated by ambient coronal plasma, may also rise to the level of GLEs, with no boost from pre-accelerated impulsive suprathermal ions. The presence of an impulsive-SEP association is neither necessary nor sufficient to produce a high-energy gradual SEP event or a GLE. It is incidental and of no apparent help at the energies considered in this study.

Several of the events seem to show somewhat lower temperatures early in the event than are seen later (see Figures 2, 4, and 8). This is common, and suggests the possibility of temperature filtration during transport especially if the source population is not isothermal. A low-temperature population of ions has systematically higher values of *A/Q* than a higher-temperature population, so these ions will be scattered less and appear preferentially among the earlier arrivals. The higher-temperature population, with lower *A/Q* will be scattered more and retarded. Even in an isothermal population, statistical variations in a distribution of *A/Q* values will also be filtered, making the early higher-*A/Q* arrivals seem cooler than the later ions with lower *A/Q*. The observed variations place a limit on the temperature spread.

Any physical mechanism which arrays ion enhancements or suppressions that can be approximated as a power law in *A/Q* provides the leverage we need in finding the source temperature that produced the best-fit pattern of *Q* values. For the impulsive SEP events this mechanism involved acceleration during magnetic reconnection (Reames, Cliver, and Kahler 2014a, b, 2015) and here it was scattering during transport of the ions out to us from a source near the Sun. This use of *A/Q* provides an entirely new method to determine ionization states of elements that differs from direct instrumental measurement (*e.g.* DiFabio *et al.* 2008), generally limited to energies below 1 MeV amu$^{-1}$, and measurements using the geomagnetic field (*e.g.* Leske *et al.* 1995, 2001). The latter measurements may differ because of additional ionization during transit, especially for ions from impulsive events which may begin fairly deep in the solar corona. Determination of a source temperature helps us identify the origin of the ions. We can now measure SEP source temperatures using observed element abundances.

**Acknowledgments**: The author thanks Steve Kahler, Lun Tan and Allan Tylka for helpful discussions.

# Source Plasma Temperatures in Gradual SEP Events

# Appendix A.  Scattering as a Power-law in A/Q?

In this paper we assume that element enhancements or suppressions caused by scattering during transport have approximately power-law dependence upon *A/Q*. How good is this approximation?

Given that the scattering mean free path $\lambda_X$ depends upon $(A_X/Q_X)^\alpha$, but is independent of distance R, we can use the expression for the solution to the diffusion equation (from Equation 5 in Tylka *et al.* 2012 or Equation C3 in Ng, Reames, and Tylka 2003 based upon Parker 1963) to write the enhancement of element X relative to O as a function of time *t* as

$$X/O = r^{-3/2} \exp \{ (1-1/r) \tau/t \} \qquad (A1)$$

where $r = \lambda_X / \lambda_O = ( (A_X/Q_X) / (A_O/Q_O) )^\alpha$ and we have redefined the parameter $\tau$, factoring the *r*-dependence from it.  In Figure A1 we plot X/O *vs.* the relative value of $(A_X/Q_X) / (A_O/Q_O)$ for several values of $\tau/t$.  The value of $\alpha$=0.6 was used in this sample. This is nearly twice the value of $\alpha$=1/3 for scattering by a Kolmogorov spectrum of waves.  The nonlinearity increases with $\alpha$.

**Figure A1.** The enhancement of an element, X/O, is shown as a function of its value of $A_X/Q_X$ relative to that of the reference, O, for several values of the time variable $\tau/t$.  A dashed fit line is plotted for comparison with the curve for $\tau/t$ = 8.

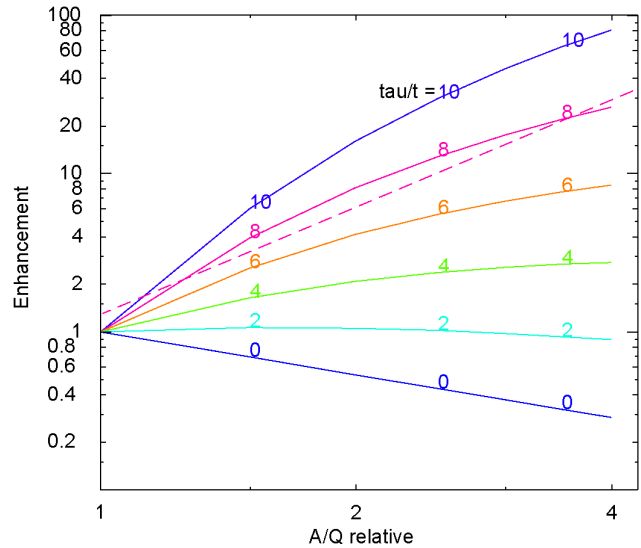

For X=Fe, experimentally observed enhancements rarely exceed 10, implying that $\tau/t \leq 8$, and *A/Q* for Fe rarely exceeds that of O by a factor of 4 (see Figure 1).  The *A/Q* dependence at late times when $\tau/t$ =0 is a power law.  A dashed fit line is plotted for comparison with the curve for $\tau/t$=8.  The discrepancy reaches ≈20% at most and varies smoothly across the range.  It cannot regroup elements with different values of $A/Q$ to compensate for the complex variation seen in Figure 1.

Formally, we can achieve a linear approximation if we remember the expansion of $\log x = (1-1/x) + (1-1/x)^2/2 + \ldots$ (for $x > \frac{1}{2}$). Using only the first term to replace $1-1/r$ with $\log r$ in Equation A1, we have

$$\mathrm{X/O} \approx r^{\,\tau/t\,-\,3/2} \tag{A2}$$

for $r > \frac{1}{2}$, as an expression for the power-law dependence of enhancements on $A/Q$ of species X. Since we can choose He or O as a reference, we can always insure that $r \geq 1$.

## Appendix B

Table B1 shows properties of the gradual SEP events for which we have been able to determine source plasma temperatures. Successive columns show the source CME onset time, the end time of SEP accumulation, the CME speed, associated flare location, GLE? (true=1), stripped ions present? (true=1), and the derived source plasma temperature.

**Table B1** Source plasma temperatures of gradual SEP events

| | Onset (UT) | SEP End (UT) | $V_{CME}$ (km s$^{-1}$) | Location | GLE | Stripping | T (MK) |
|---|---|---|---|---|---|---|---|
| 1 | 97 Nov  4 0526 | 97 Nov  6 0000 | 785 | S14  W33 | 0 | 0 | 3.2±0.3 |
| 2 | 97 Nov  6 1137 | 97 Nov  9 0000 | 1556 | S18  W63 | 1 | 0 | 2.5±0.2 |
| 3 | 98 Apr 20 0956 | 98 Apr 24 0000 | 1863 | S   W90 | 0 | 1 | ≤0.8+0.2 |
| 4 | 98 Aug 24 2212 | 98 Aug 27 0800 | - | N35  E09 | 1 | 0 | 1.6±0.2 |
| 5 | 98 Sep 30 1350 | 98 Oct  2 0800 | - | N19  W85 | 0 | 0 | 1.3±0.2 |
| 6 | 98 Nov 14 0518 | 98 Nov 16 0800 | - | N   W120 | 0 | 0 | 2.9±0.4 |
| 7 | 00 Jul 14 1025 | 00 Jul 18 0800 | 1674 | N22  W07 | 1 | 0 | 1.16±0.2 |
| 8 | 00 Sep 12 1145 | 00 Sep 14 1600 | 1550 | S17  W09 | 0 | 0 | 1.6±0.2 |
| 9 | 00 Nov  8 2250 | 00 Nov 11 0800 | 1345 | N10  W75 | 0 | 0 | 1.16±0.2 |
| 10 | 01 Apr  2 2143 | 01 Apr  5 1600 | 2505 | N17  W78 | 0 | 0 | 3.0±0.3 |
| 11 | 01 Apr 15 1332 | 01 Apr 17 0800 | 1199 | S20  W84 | 1 | 0 | 3.1±0.4 |
| 12 | 01 Aug 15 2337 | 01 Aug 17 1600 | 1575 | ?  W140 | 0 | 0 | 1.4±0.4 |

| 13 | 01 Sep 24 1021 | 01 Sep 29 1600 | 2402 | S16 E23 | 0 | 0 | 0.9 ±0.3 |
|----|----------------|----------------|------|---------|---|---|----------|
| 14 | 01 Oct  1 0529 | 01 Oct 3 1600 | 1405 | S20 W88 | 0 | 0 | 1.9±0.2 |
| 15 | 01 Nov 4 1612 | 01 Nov 5 1800 | 1810 | N06 W18 | 1 | 0 | 0.9±0.3 |
| 16 | 01 Dec 26 0506 | 01 Dec 28 0000 | 1406 | N08 W54 | 1 | 0 | 1.6±0.2 |
| 17 | 02 Apr 21 0117 | 02 Apr 24 0800 | 2409 | S14 W84 | 0 | 0 | ≤0.8+0.3 |
| 18 | 02 May 22 0323 | 02 May 24 0000 | 1494 | S22 W53 | 0 | 0 | 1.26±0.2 |
| 19 | 02 Jul 15 1950 | 02 Jul 18 0800 | 1132 | N19 W01 | 0 | 1 | 1.2±0.2 |
| 20 | 03 Oct 26 1741 | 03 Oct 28 0000 | 1537 | N02 W38 | 0 | 0 | 1.6±0.3 |
| 21 | 03 Oct 28 1106 | 03 Oct 29 0000 | 2459 | S16 E08 | 1 | 0 | ≤0.8+0.2 |
| 22 | 03 Oct 29 2041 | 03 Oct 31 1600 | 2029 | S15 W02 | 1 | 0 | 2.5±0.4 |
| 23 | 03 Nov 2 1720 | 03 Nov 4 1600 | 2598 | S14 W56 | 1 | 0 | 1.0±0.3 |
| 24 | 03 Dec 2 1024 | 03 Dec 4 0000 | 1393 | S14 W70 | 0 | 1 | 1.3±0.2 |
| 25 | 04 Apr 11 0400 | 04 Apr 12 1600 | 1645 | S16 W46 | 0 | 1 | 1.26±0.2 |
| 26 | 04 Jul 25 1441 | 04 Jul 28 0000 | 1333 | N08 W33 | 0 | 0 | 1.2±0.2 |
| 27 | 04 Sep 13 0031 | 04 Sep 15 0800 | 1328 | N03 E49 | 0 | 1 | 1.4 ±0.2 |
| 28 | 04 Nov 7 1622 | 04 Nov 9 1600 | 1759 | N09 W17 | 0 | 1 | 0.9±0.2 |
| 29 | 05 Jan 15 2240 | 05 Jan 17 1600 | 2861 | N15 W05 | 0 | 1 | 1.0±0.4 |
| 30 | 05 May 13 1648 | 05 May 15 0800 | 1689 | N12 E11 | 0 | 1 | 1.6±0.4 |
| 31 | 05 Jul 14 1027 | 05 Jul 17 0000 | 2115 | N11 W90 | 0 | 0 | ≤0.8+0.2 |
| 32 | 05 Aug 22 1705 | 05 Aug 25 0000 | 2378 | S13 W65 | 0 | 1 | 1.0±0.2 |
| 33 | 05 Sep 13 1942 | 05 Sep 15 1600 | 1866 | S09 E10 | 0 | 0 | 1.4±0.2 |
| 34 | 06 Dec 13 0225 | 06 Dec 14 1600 | 1774 | S06 W26 | 1 | 0 | 2.0±0.2 |
| 35 | 11 Aug 4 0339 | 11 Aug 6 0800 | 1315 | N19 W36 | 0 | 0 | 3.2±0.5 |
| 36 | 12 Jan 23 0346 | 12 Jan 26 1600 | 2175 | N28 W21 | 0 | 1 | 1.6±0.2 |
| 37 | 12 Jul 6 2255 | 12 Jul 8 1200 | 1828 | S13 W59 | 0 | 0 | 2.5±0.3 |
| 38 | 13 Apr 11 0650 | 13 Apr 13 1200 | 0861 | N09 E12 | 0 | 0 | 2.0±0.3 |
| 39 | 13 May 22 1255 | 13 May 26 0000 | 1466 | N15 W70 | 0 | 0 | 1.0±0.4 |
| 40 | 13 Sep 29 2152 | 13 Oct 2 0000 | 1179 | N10 W33 | 0 | 1 | 1.4±0.4 |
| 41 | 13 Oct 28 0417 | 13 Nov 1 0000 | 1201 | N08 W71 | 0 | 0 | 2.8±0.4 |
| 42 | 14 Jan 6 0733 | 14 Jan 7 1200 | 1402 | S04 E13 | 0 | 0 | 2.9±0.4 |
| 43 | 14 Jan 7 1805 | 14 Jan 12 0000 | 1830 | S15 W11 | 0 | 0 | 1.2±0.5 |

Source Plasma Temperatures in Gradual SEP Events

| | | | | | | | | |
|---|---|---|---|---|---|---|---|---|
| 44 | 14 Apr 18 1243 | 14 Apr 21 0000 | 1203 | S20 | W34 | 0 | 0 | 2.8±0.3 |
| 45 | 14 Sep 10 1728 | 14 Sep 17 0000 | 1767 | N14 | E02 | 0 | 0 | 0.9±0.3 |